\documentclass[doublecol]{epl2}
\usepackage{graphicx}

\newcommand{\e}{\mathrm{e}}

\newcommand{\kbt}{k_{\mathrm{B}}T}
\newcommand{\lb}{l_{\mathrm{B}}}
\newcommand{\lgc}{\lambda_\mathrm{GC}}
\newcommand{\ld}{\lambda_\mathrm{D}}
\newcommand{\lpm}{\lambda_{\pm}}
\newcommand{\lp}{\lambda_{+}}
\newcommand{\lm}{\lambda_{-}}

\newcommand{\df}{\mathrm{d}}

\begin{document}

\title{Electrostatic Interactions of Asymmetrically Charged Membranes}
\author{Dan Ben-Yaakov\inst{1}, Yoram Burak\inst{2}, David Andelman\inst{1,3} \and S. A. Safran\inst{3}}

\institute{
  \inst{1} Raymond and Beverly Sackler Faculty of Exact Sciences,
  Tel Aviv University, Ramat Aviv, Tel Aviv 69978, Israel\\
  \inst{2} Kavli Institute for Theoretical Physics, University of California, Santa Barbara, CA 93106 USA\\
  \inst{3} Department of Materials and Surfaces, Weizmann Institute
  of Science, Rehovot 76100, Israel}


\abstract
{We predict the nature (attractive or repulsive) and range
(exponentially screened or long-range power law) of the
electrostatic interactions of oppositely charged and planar plates
as a function of the salt concentration and surface charge densities
(whose absolute magnitudes are not necessarily equal). An
analytical expression for the crossover between attractive and
repulsive pressure is obtained as a function of the salt
concentration. This condition reduces to the high-salt limit of
Parsegian and Gingell where the interaction is  exponentially
screened and to the zero salt limit of Lau and Pincus in which
the important length scales are the inter-plate separation and the
Gouy-Chapman length. In the regime of low salt and high surface
charges we predict –-- for any ratio of the charges on the surfaces ---
that the attractive pressure is long-ranged as a function of the
spacing.
The attractive
pressure is related to the decrease in counter-ion
concentration as the inter-plate distance is decreased. Our theory predicts
several scaling regimes with different scaling expressions for the
pressure as function of salinity and surface charge densities.
The  pressure predictions can be related to surface force
experiments of oppositely charged surfaces that are prepared
by coating one of the mica surfaces
with an oppositely charged polyelectrolyte.}

 \maketitle


\section{Introduction}
The interactions between oppositely charged surfaces are important
in both biological and materials science contexts.  Proteins contain
both cationic and anionic regions and in some cases, interactions
between proteins are due to unlike charge attraction, mediated by
the intervening counterions and salt.  The delivery of cationic
vesicles to cells –-- of interest in gene therapy applications --–
involves counterion and salt mediated electrostatic interactions of
two oppositely charged membranes~\cite{safinya1,benshaul1,bruinsma}.
Similar considerations may also be important in understanding adhesion
and fusion in systems of oppositely charged bilayers~\cite{sackmann,hed}.  Recent
experiments on hydrophobically prepared mica surfaces have indicated
that such surfaces have domains with different charges and the
observed long-range attractions (in the nanometer regime) may again
be related to unlike charge attractions mediated by counterions and
salt \cite{klein1,isra2}.

In this paper we predict the interactions between two homogeneously
charged surfaces with opposite charge.  The surfaces are in aqueous
solution that contains the counterions and added salt.  The  interactions
can be attractive or repulsive,  short-ranged (exponentially decaying)
or long-ranged (power-law)
depending on the ratio of the distance between
the surfaces to the important length scales of the problem:
(i) The Gouy-Chapman length,
$\lgc=1/(2\pi\lb\sigma)$ that is inversely proportional to the surface
charge density $\sigma$;
(ii) the Bjerrum length $\lb=e^2/\varepsilon\kbt$
equal to about 7\,\AA\ in water ($\varepsilon\simeq 80$) at room
temperature; and, (iii) the Debye-H\"uckel (DH) length, $\ld=1/\sqrt{8\pi\lb c_b}$
where $c_b$ is the bulk 1:1 salt concentration.  Since there are several
length scales there are several regimes that characterize the interactions.

In order to model interactions between charged surfaces, membranes
and particles, one typically considers two planar
surfaces separated by a distance, $d$. Previous studies considered
the {\em symmetric} case where the two surfaces have fixed and
equal surface charge densities. Within Poisson-Boltzmann (PB) theory, it can be
shown that the interaction between two surfaces with the same charge is always
repulsive due to the counter-ion entropy. Other refinements include
correction to the PB theory  especially in the limit of strong
surface charges and multi-valent counterions.  The theory of the
interactions between two surfaces with opposite charges
has received less attention. The pioneering study of Parsegian and Gingell
\cite{parsegian1} considered high salt concentrations and used  the
linear DH theory, to predict regions of repulsive and attractive
inter-plate pressure as a function of the distance and salt
concentration. Lau and Pincus~\cite{pincus1} considered the case of
two oppositely charged surfaces with no added salt and found a
simple analytical criterion for the crossover between attraction and
repulsion.

The interesting, intermediate region of high surface charge and
relatively low salt concentrations requires the use of the
non-linear PB theory.  There can be a large regime in which the
distance between the plates is smaller than the DH screening length
(which can be very large at low salt concentrations) but larger than
the Gouy-Chapman length (which can be very small for high surface
charge) in which the non-linear treatment must be applied.  The case
of two plates with equal and opposite surface charges was considered
recently by several authors~\cite{safran2,grunberg1}. In this case,
where the two plates are electrically neutral and the ions in
solution have no net charge, the force is always attractive. At some
characteristic inter-plate distance, $d^*$, which scales as
$\lgc\ln(\ld/\lgc)$, the counterions are released into the bulk
reservoir provided that $\lgc\ll d^*\ll\ld$. In addition, for
inter-plate separation in the range of intermediate values (between $\ld$ and
$\lgc$), the pressure was shown to be long-ranged, and scales with
the inverse of $d^2$~\cite{safran2}. In the present work, we
generalize these results and consider two oppositely charged
surfaces whose charges are not necessarily equal in magnitude.  This
general case is important in order to analyze experiments in which
surfaces are not completely antisymmetric. We use our analytical
theory to predict  a crossover between attraction and repulsion and
the counterion release concept is extended to the asymmetric case.
We also are able to consider several limiting regimes for the
asymmetric case that complement the numerical solutions of the
problem.

\section{Poisson Boltzmann model}
The Poisson-Boltzmann (PB) theory is a mean-field theory that relates the
electric potential,
$\psi(\vec{r})$, and the Boltzmann distribution for the ion number
density, $c(\vec{r})$, at thermodynamic equilibrium.
For two surfaces  immersed in a 1:1
monovalent ionic solution, and for a dimensionless potential
$\phi\equiv e\psi/k_\mathrm{B}T$, the PB equation reads:
\begin{equation}
 \nabla^2\phi=\kappa_\mathrm{D}^{2}\sinh\phi=\ld^{-2}\sinh\phi\,
 ,
\label{pbsinheq}
\end{equation}
where $\ld=\kappa_\mathrm{D}^{-1}$ is the Debye-H\"uckel (DH) screening
length defined above.

 We  consider two charged surfaces that are infinite in extent in the
$(x,y)$ plane and are separated in the $z$ direction
by a distance  $d$. In this case, the PB
equation reduces to a one-dimensional equation in the coordinate $z$ perpendicular to
the planes.
The focus of this paper is on two
oppositely charged plates that we call the  {\em asymmetric two
plate} problem. A positively charged plate with charge density
$\sigma_{+}>0$ is located at $z=d/2$, and a negatively charged one
with $\sigma_{-}<0$ is located at $z=-d/2$. While in the
well-studied symmetric case, where  $\sigma_{+}=\sigma_{-}$,
\cite{andelman1} the potential (and ion densities) are symmetric
about the  midplane $z=0$, in the asymmetric case, the midplane is
no longer a plane of symmetry.  Instead, a separate boundary
condition at each plate must be explicitly considered. These
conditions relate the electric field at each plate to the surface
charge density:
\begin{equation}
\label{boundaries}
 \frac{\df\phi}{\df z}\bigg |_{z=\pm d/2}
 = {4\pi \lb|\sigma_{\pm}|} \equiv \frac{2}{\lpm}>0\,
\end{equation}
where $\lpm$ are the Gouy-Chapman lengths for the corresponding
surfaces.

The spatial dependence of the potential and ion densities is
obtained by solving the PB equation, eq.~(\ref{pbsinheq}) subject to
the boundary conditions, eq.~(\ref{boundaries}). The profiles
predict the local concentration of the mobile ions and their
associated potentials; these can be measured, for example, using
scattering techniques. However, more often, the forces exerted on
the charged plates are measured \cite{isra1}. For a given separation
$d$,  the pressure (or equivalently, the force per unit area) must
be constant in the entire region between the plates if the system is
in thermodynamic equilibrium. Thus, it is only necessary to calculate
the pressure at any convenient point, $z$, within the gap. Because
the two-plate system is in contact with a reservoir of mobile ions
(the salt reservoir), the net pressure (in units of $\kbt$) exerted
on the plates is given by the difference between the inner and outer
pressures, $\Pi=P_{\rm in}-P_{\rm out}$.  This pressure can be
calculated, for example, by integrating  the PB equation once and
relating the integration constant with the pressure $\Pi$ \cite{andelman1}:
\begin{equation}
\label{p_total} \Pi=-\frac{1}{8\pi\lb}\left(\frac{\df \phi}{\df
z}\right) ^2+2c_b(\cosh\phi-1)\, .
\end{equation}
The $z$-independent osmotic pressure
comprises  two terms: (i) an
attractive contribution whose origin is the electrostatic energy;
this has the form of a negative term proportional to the square of
the electric field \cite{landau1} and (ii) a repulsive contribution
that arises from the translational entropy of the ions and is given
by the ideal-gas law. In this term we have already subtracted
the outer pressure, $P_{\rm out}=2c_b $ as is explained above.

It is easy to show that the pressure for the case of symmetrically
charged plates is always repulsive ($\Pi>0$) within the PB
approximation. However, in the general, asymmetric case the pressure
can be either repulsive ($\Pi>0$) or attractive ($\Pi<0$)  as we
explain below.

Our two-plate problem is fully determined by four physical
parameters: The two surface charge densities $\sigma_\pm$, the ionic
strength, $c_b$, and  the separation $d$. However, by using normalized
variables, it is easy to show that the problem is uniquely defined
by three ratios: $\lpm/\ld$ and $d/\ld$, where $\lpm$ is related to
$\sigma_\pm$ in eq.~(\ref{boundaries}) and $\ld$ is related to
$c_b$. In our PB solution we use an alternative parametrization
scheme including the dimensionless pressure $\hat{\Pi}=\Pi/c_b$, the surface
potentials $\phi_\pm=\phi(\pm d/2)$. We can now relate those three
parameters to the original ones by three relations. Two relations
can be derived from eq.~(\ref{p_total}):
\begin{equation}
 \hat{\Pi}=2\left(\cosh\phi_{\pm}-1\right)-\left(\frac{\ld}{\lpm}\right)^2\, .
\label{surfacephi}
\end{equation}
Integration
between the two boundaries, $\pm d/2$, gives an additional, third
relation:
\begin{equation}
 \frac{d}{\ld}=\int_{\phi_{-}}^{\phi_{+}}
 \frac{\df\phi\,'}{\sqrt{2\left(\cosh\phi\,'-1\right)-\hat{\Pi}}}\,
 .
\label{elliptic2h}
\end{equation}
We can now express the PB solution and pressure $\hat{\Pi}$ [via
eqs.~(\ref{surfacephi}) and (\ref{elliptic2h})]  as a function of
the parameters that characterize the physical system – namely, the
surface charge densities, $\lp,\lm$, the DH length, $\ld$
(determined by the salt concentration), and the separation between
the plates, $d$. For a few simplified cases, an analytical solution
exists, while in the general asymmetric case we can write the
potential, $\phi$, in terms of elliptic integrals \cite{borkovec}
whose solution can be obtained only numerically. We next present a
general analytical result that predicts when the interaction crosses
over from repulsive to attractive as a function of the system
parameters. The physical origin of this crossover is the competition
between the electrostatic and entropic interactions described above.

\section{Attractive to repulsive crossover}

The condition $\Pi=0$ in eq.~(\ref{p_total}) for the asymmetric, two plate system
determines the cross-over from
repulsive to attractive interactions in the system:
\begin{equation}
\label{single_pb_eq} -\frac{1}{8\pi\lb}(\phi^\prime)^2 + 2c_b (\cosh\phi-1)=0\, .
\end{equation}
This is a relation between the potential, $\phi(z)$, and its
derivative $\phi^\prime$ at any point $z$. The condition, $\Pi=0$,
also fixes one relation between the three dimensionless ratios:
$\lpm/\ld$ and $d/\ld$. Namely, this confines the system to a two
dimensional surface in the three-dimensional parameter space
($\lp/\ld$, $\lm/\ld$, $d/\ld$).

An analytical expression for this crossover is found by observing
that the case of $\Pi=0$ can be exactly mapped onto the equations
that describe a single plate
in contact with the same reservoir \cite{burak1}. For this purpose, we consider a
system with a single, positively charged plate at $z=0$.
The analytical expression for the potential and its derivative
(electric field) are well known
\begin{eqnarray}
\label{single_pot}
\phi&=&2\ln\left(\frac{1+\gamma_+\e^{-z/\ld}}{1-\gamma_+\e^{-z/\ld}}\right)\\
 \phi^\prime &=&-\frac{1}{\ld}\frac{4
\gamma_+\e^{-z/\ld}}{1-\gamma_+^2\e^{-2z/\ld}} \label{single_elec_field}
\end{eqnarray}
where
$\gamma_{+}=\sqrt{\left(\lp/\ld\right)^2+1}-\left(\lp/\ld\right)$.
The mapping between the two problems is simply done by requiring
that at distance $d$ away from the $z=0$ charged plate (the single
plate case) the electric field is equal to the electric field as
determined from Gauss's law at the negative plate located at $z=-d/2$ in the
two-plate problem: $\phi^\prime(d)=-2/\lm$. This results in the
relation:
\begin{equation}\label{gamma_pm}
\gamma_{+}=\e^{d/\ld}\,\gamma_{-}
\end{equation}
where $\gamma_{-}$ is similarly defined as
$\gamma_{-}=\sqrt{\left(\lm/\ld\right)^2+1}-\left(\lm/\ld\right)$.

\begin{figure}[!h]\centering
\includegraphics[width=0.35\textwidth]{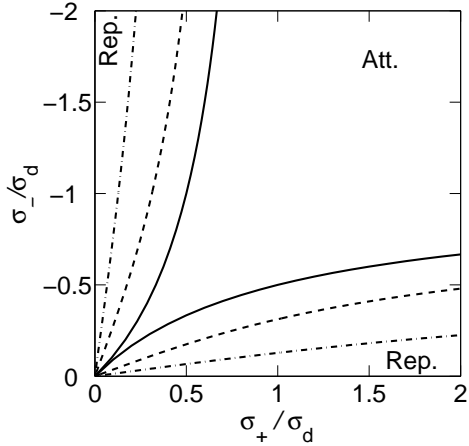}
\caption{Regions of attraction and repulsion. $\sigma_{+}$ and
$\sigma_{-}$ are the surface charge densities, and
$\sigma_{d}=1/2\pi l_Bd$. The dashed and dash-dot lines represent
the exact expression [eq.~(\ref{cond_att_rep})] for crossover from
attractive to repulsive interactions with
$d/\ld=1$ and 2, respectively. The solid line shows the no-salt
limit [eq.~(\ref{no_salt_cond})]. } \label{attraction}
\end{figure}

The  relation~(\ref{gamma_pm}) between $\gamma_\pm$ (or $\lpm$) is
equivalent to the $\Pi=0$ crossover in the asymmetric two plate
system, between the repulsive, $\Pi>0$, and attractive, $\Pi<0$,
regimes. When $\gamma_-\rightarrow 0$ (the negative plate is
neutral, $\sigma_{-}\rightarrow 0$), the plates must repel each
other. In addition, although we have only  treated so far the case
$\sigma_{+}>|\sigma_{-}|$, our results are quite general since the
two-plate pressure is invariant under the exchange $\sigma_{+}
\leftrightarrow |\sigma_{-}|$. Therefore, the condition for
attraction reads:
\begin{equation}
\label{cond_att_rep}
\e^{-d/\ld}<\frac{\gamma_+}{\gamma_-}<\e^{d/\ld},
\end{equation}
This result is plotted on fig.~\ref{attraction} where two lines
separate a central region of attractive interactions from two wedges
in the ($\sigma_{-}$, $\sigma_{+}$) plane, that denote repulsive
interactions. In our plots, the charge densities are normalized by
$\sigma_D=1/2\pi\lb d$.

\begin{figure}[!h]\centering
\includegraphics[width=0.35\textwidth]{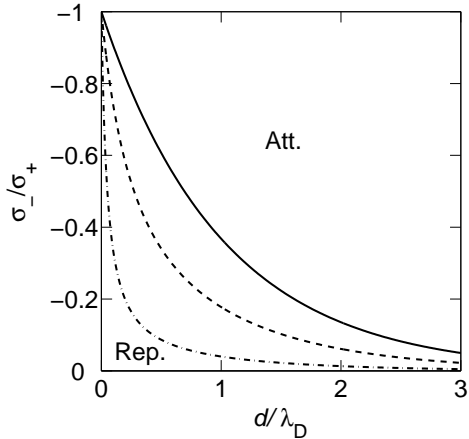}
\caption{The zero-pressure line $\Pi(\sigma_{-}/\sigma_{+},d)=0$ in
the ($\sigma_{-}/\sigma_{+},d/\ld$) plane. The dashed line
represents the exact expression [eq.~(\ref{cond_att_rep})] for
crossover with $\lp/\ld=0.3$, corresponding to $c_b\simeq10^{-4}\,$M
for typical mica effective net surface charge, $\sigma_+\simeq
e/4000\,$\AA$^2$. On the dash-dot line  $\lp/\ld=0.05$, where for
the same  $c_b\simeq10^{-4}\,$M the surface charge is
$\sigma_{+}\simeq e/670\,$\AA$^2$. The crossover in the DH limit
[eq.~(\ref{highsaltattr1})] is shown in solid line.}
\label{attraction_DH}
\end{figure}

This result  is exact for arbitrary salt concentration and surface
charge densities. It has two limits that have been previously studied.
One limit, that  of zero salt, was
analyzed by Lau and Pincus~\cite{pincus1}. In this limit, the
counterions in the solution balance the surface excess charge
$\Delta\sigma=\sigma_{+}-|\sigma_{-}|$. Formally, we obtain this
limit by taking $\ld\to\infty$ in eq.~(\ref{cond_att_rep}).
Expanding $\gamma_\pm$  in powers of $\lpm/\ld\,$,
$\gamma_{\pm}\simeq 1-\lpm/\ld$, and eq.~(\ref{cond_att_rep})
yields:
\begin{equation}
\label{no_salt_cond}
\left|\frac{1}{\sigma_{+}}-\frac{1}{|\sigma_{-}|}\right|<\frac{1}{\sigma_{d}}\,
,
\end{equation}
The crossovers from attraction to repulsion  are plotted in
fig.~\ref{attraction} for several salt concentrations using eq.~(\ref{cond_att_rep})
and the no-salt limit, eq.~(\ref{no_salt_cond}).
The second, well-known limit of high salt concentrations,
$\ld\ll\lpm$, was studied by Parsegian and Gingell \cite{parsegian1}
in the 1970s by linearizing the PB equation. They  derived the
pressure in the linear DH regime. In this case,  $\gamma_\pm$ can be approximated by
$\gamma_\pm\simeq\ld/2\lpm$ in
eq.~(\ref{cond_att_rep}), and the attraction condition is
\begin{equation}
\label{highsaltattr1}
\e^{-d/\ld}<\frac{\sigma_{+}}{|\sigma_{-}|}<\e^{d/\ld}
\end{equation}
which reproduces the result of Parsegian and Gingell. In
fig.~\ref{attraction_DH} the line of zero pressure that separates
attractive from repulsive interactions is plotted in the
($\sigma_{-}/\sigma_{+},d/\ld$) plane for three salt concentrations.
The plot shows the high salt limit of Parsegian and Gingell [from
eq.~(\ref{highsaltattr1})], as well as an intermediate amount of
salt ($\lp/\ld=0.3$), and a relatively low amount of salt
($\lp/\ld=0.05$) from eq.~(\ref{cond_att_rep}).

We now compare the two figures and comment on the role of salt.
In fig.~1 it is apparent that increasing the salt concentration
enlarges the attractive region at the expense of the repulsive one.
Another remark is that for $|\sigma_\pm|>\sigma_d$ (large $d$ and/or
strongly charged plates), the pressure is always negative
(attraction) with no dependence on other system parameters like
$\sigma_{-}/\sigma_{+}$ and $\ld$. This is related to the asymptotic
behavior of the no-salt crossover curves.
\footnote{Another observation is
that in fig.~2 it looks like the predictive behavior is
opposite as compared to fig.~1 because for larger amounts of salt
the repulsive region is enlarged. However, the two figures are in
total accord because in fig.~2 the $x$-axis is also scaled by $\ld$.}

\section{Gouy-Chapman pressure}

For the exact anti-symmetric case ($\sigma_{+}$=$-\sigma_{-}$)
and in the regime of low salt and strongly charged plates it was
shown~\cite{safran2} that an approximate scaling relation between
the pressure $\Pi$ and $d$ is: $\sqrt{{c_b}/{|\Pi|}}
\ln(|\Pi|/{c_b})\sim {d}/{\ld}$.  The range of
validity of this scaling relation is the diagonal ray in fig.~3,
$\sigma_{+}/\sigma_d=|\sigma_{-}|/\sigma_d\gg 1$, and in addition
$d\ll \ld$. This result can be extended to the more general
asymmetric region bounded by the hashed box in fig.~3: $\sigma_{+}/\sigma_d\gg 1$ and
$|\sigma_{-}|/\sigma_d\gg 1$. From the integral in
eq.~(\ref{elliptic2h}) the relation between the pressure and $d$ is
deduced to be
\begin{equation}
 2\sqrt{\frac{c_b}{|\Pi|}}
\ln\left(\frac{4|\Pi|}{c_b}\right)\simeq \frac{d}{\ld}\left(1+
\frac{\sigma_d}{\sigma_{+}}+\frac{\sigma_d}{|\sigma_{-}|}\right)\, .
\label{exact2}
\end{equation}
%

\begin{figure}[!h]\centering
\includegraphics[width=0.28\textwidth]{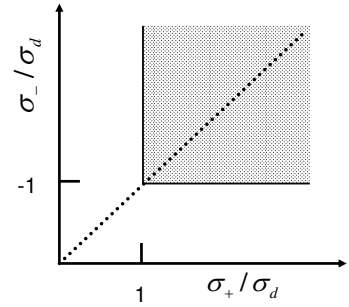}
\caption{The region of validity of the scaling relation is denoted
by the hashed box in the ($\sigma_{+}/\sigma_d$,
$\sigma_{-}/\sigma_d$) plane. The dashed line denotes the
anti-symmetric case.}
\end{figure}

We note that in the entire range of validity of this scaling
expression, the right hand side of eq.~(\ref{exact2}) varies between ${d}/{\ld}$ and $3d/\ld$, meaning
that it is roughly described by the anti-symmetric result mentioned
above \cite{safran2}. Thus, as long as the plates are strongly
charged, $\Pi$ does not depend on the surface charge densities.
This relation can be solved iteratively for $\Pi(d)$ and the first
iteration yields:
\begin{equation}\label{P_SAF1}
\Pi\simeq-\frac{2}{\pi\lb d^2}\ln^2\left(d/8\ld\right)\, .
\end{equation}
%

\section{Counter-ion release}

In the limit of infinite separation between the plates, an appropriate
concentration of mobile cations and
anions accumulate in the vicinity of each plate in order to
neutralize the surface charge. When the plates are brought closer
these two, oppositely charged clouds of mobile ions begin to overlap. Pairs of
negative and positive counterions can thus neutralize each other and escape
to the reservoir, where they gain entropy with no cost of electrostatic
energy. This phenomenon of counterion release is the physical origin of the
attractive forces between oppositely charged plates
\cite{safran2,grunberg1}. The parameter that characterizes the release of the
counterions is defined as
the excess charge per unit volume integrated over the entire
separation between the surfaces:in
eq.~(\ref{cond_att_rep}),
\begin{equation}
\label{eta_int}
 \eta\equiv\int_{-d/2}^{+d/2}\df z\left[c_+(z)+c_-(z)-2c_b\right].
\end{equation}
where $c_\pm(z)$ are the number densities of the cations and anions.
Safran \cite{safran2} considered the relation between the pressure
and the fraction of ions released into the reservoir for the exact
antisymmetric case, ($\sigma_{+}$=$-\sigma_{-}$), in the regime of low
salt. Here we present a generalization of this result to the
asymmetric case, where we focus on the same low salt, high
surface charge regime but with
$\Delta\sigma=\sigma_{+}-|\sigma_{-}|$ which is small compared to $\sigma_{\pm}\,$. Under
these assumptions the integral in eq.~(\ref{eta_int}) can be
evaluated and gives:
\begin{equation}
 \eta\simeq2\sigma_{+}\left(1-\sqrt{2\pi\lb\lp^2|\Pi|}\,
 \right)-\Delta\sigma\, ,
\label{finaleta}
\end{equation}
In the limit of $\Pi\rightarrow 0$ we find
$\eta\simeq\sigma_{+}+|\sigma_{-}|$; this represents the largest
possible value of excess counterions (beyond the bulk value
$\sigma_b=2dc_b$). In this limit, corresponding to large separations
[eq.~(\ref{P_SAF1})] each of the surfaces is neutralized by its own
cloud of counterions. As the separation between the two surfaces
decreases, the pressure reaches its maximal attractive value
$|\Pi|\rightarrow1/2\pi\lb\lm^2$, and the counterion excess
approaches its minimal value: $\eta\simeq\Delta\sigma$. In this
situation, not all the counterions are forced to remain in the gap
between the plates. The two oppositely charged plates screen each
other except for an excess of surface charge $\Delta\sigma$ that is
compensated by the remaining counterions.

The largest attractive pressure occurs for a distance $d^*$ that
can be estimated as $d^*=2\lm\ln{2\ld/\lm}\,$. 
Note the difference
in this situation between the asymmetric and the exact
anti-symmetric case for which the 
pressure saturates at $d<d^*$.

\section{Scaling regimes}

In the general asymmetric case we must consider the charge asymmetry
ratio, $\lp/\lm$ (or $|\sigma_{-}|/\sigma_{+}$) as an additional
parameter. We investigate several scaling regimes in the
three-dimensional parameter space: ($\lp/\ld$, $\lp/\lm\,$,
$d/\ld$). The pressure, as a function of the asymmetry ratio, can be treated
by considering one of the following two limits: {\rm i}) in the limit of
$|\sigma_{-}|/\sigma_{+}\ll 1$, the negative plate can be taken as a
neutral one, implying a repulsion between the plates, as
demonstrated by eq.~(\ref{cond_att_rep}). In this limit, there is a
mathematical correspondence between the asymmetric problem and the
symmetric one ($\sigma_{+}=\sigma_{-}$) with about twice the surface
separation, $d\rightarrow 2d\,$. The symmetric configuration
satisfies the boundary conditions corresponding to the case of one
neutral and one charged surface by the vanishing of the electric
field at the mid-plane. In this symmetric-like limit the pressure
scales like the pressure in the symmetric case, as discussed in
detail at ref.~\cite{andelman1}.
\begin{figure}[!h]\centering
\includegraphics[width=0.28\textwidth]{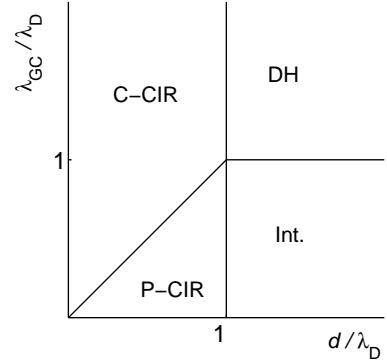}
\caption{A schematic view of the various limits of the PB equation
for two oppositely charged plates (the anti-symmetric case). The
four regimes discussed in the text are: Debye-H\"uckel (DH),
Intermediate (Int.), Partial and Complete Counter-Ion Release (P-CIR
and C-CIR). They are separated by four crossover lines:
Int.$\leftrightarrow$DH at $\lgc=\ld$ and $\ld < d$,
DH$\leftrightarrow$C-CIR at $\ld=d$ and $\ld < \lgc$,
C-CIR$\leftrightarrow$P-CIR at $d=\lgc$ and $\ld>d$,
P-CIR$\leftrightarrow$Int. at $\ld=d$ and $\lgc<\ld$.}
\label{regimes_anti_sym}
\end{figure}

{\rm ii}) On the other hand, when the surface charge densities are
nearly equal (and opposite), {\em i.e.}
$\Delta\sigma\ll|\sigma_\pm|\,$, the pressure is attractive in a
wide range of separations, and the formulae for the pressure are
similar to those for the exact, antisymmetric case. Here, for
simplicity, we present the regimes of the exact antisymmetric case
with two oppositely charged plates, $\pm\sigma$. The different
regimes are shown in the ($\lgc/\ld$, $d/\ld$) plane
[fig.~\ref{regimes_anti_sym}], and discussed below: ({\rm a})
{\em Debye-H\"uckel}. The limit of $\ld\ll \lgc$ corresponds to
low potentials ($\phi\ll1$) for which the PB equation can be
linearized. We obtain an attractive pressure ($\Pi<0$)  as expected
for two oppositely charged plates:
\begin{equation}
\Pi\simeq-\frac{2}{\pi\lb\lgc^2}\,\e^{-d/\ld}\, .
\end{equation}
This pressure expression decays exponentially
with distance.
({\rm b}) {\em Complete Counter-ion Release}. For small separations,
$d\ll \ld$ and $d \ll \lgc$, the charge neutrality of the system is
maintained by the surfaces, and all the cations and anions are
released to the reservoir. This yields a direct electrostatic
interaction of two capacitor plates of charge $\pm\sigma$ in a
dielectric medium:
\begin{equation}
\Pi\simeq-\frac{1}{2\pi\lb\lgc^2}\, .
\end{equation}
({\rm c}) {\em Partial Counterion Release}. This regime is
defined by $\lgc\ll d \ll \ld$, where the plates are strongly charged
and the salt concentration is low.  Equation.~(\ref{P_SAF1}) can be used to
predict the pressure as a function of the separation, $d$. It is
interesting to note that in this regime, the Gouy--Chapman pressure shows
a long-range, power law dependence on $d$. It is also independent
of the value of  the surface charge, similarly to the result of
Gouy-Chapman regime in the symmetric case.
({\rm d}) {\em Intermediate}. When the plates are strongly charged,
$\lgc\ll \ld$, the PB equation cannot be linearized. However, if the
separation  is large, $d\gg\ld$, the surfaces are weakly
interacting and can be treated as two separated plates. As a result,
the electric field at the midplane is given by summing these two
contributions. Under these assumptions, the electric field can be
approximated by that of a single plate, and the pressure is
\begin{equation}
\label{P_int_anti_symm1}
\Pi\simeq-\frac{8}{\pi\lb\ld^2}\e^{-d/\ld}\, .
\end{equation}

In the DH and intermediate regimes the pressure decays exponentially
with the scaled separation, $d/\ld\,$ due to screening effect of the
salt ions. In the symmetric-like limit, the characteristic decay
length is half as small. In the Gouy-Chapman regime the pressure is
independent of the surface charge density, both in the antisymmetric
and symmetric-like limits. This reveals a special
property of the electrostatic interaction at medium-range
separations: the length scales related to the surface charges,
$\lpm$ and $\ld$, have no effect on the pressure (beside a small logarithmic
correction as in eq.~(\ref{P_SAF1})). At small separations,
the boundary conditions have the largest effect on the pressure
between the plates. In the symmetric-like limit, the confinement of
ions in between the plates results in a divergence of the pressure,
whereas in the anti-symmetric like limit, the pressure saturates due
to the complete release of all the counterions at small separations.

The similarity between the symmetric and
anti-symmetric cases can be seen at large and intermediate
separations, while the distinction is evident at small separations.
It is important to note that there is one major difference between these two
limits: the symmetric-limit is purely repulsive, while the
anti-symmetric one is always attractive.

\section{Discussion}
The theoretical investigation in this work suggests that the
asymmetric Poisson-Boltzmann model  predicts several interesting 
physical regimes in
the interaction of dissimilar charged bodies immersed in electrolyte
solutions. In particular, the pressure dependence  on the separation
$\Pi(d)$ can be evaluated  from  both the scaling relations and the
numerical solutions. These results can be tested in various
experiments measuring forces between charged objects. One of the interesting
results to be tested experimentally is the
prediction for the crossover between attractive
and repulsive interactions, $\Pi=0$,
at high and low salt conditions. Others may include
the crossover of the interactions from exponential decay to power law due to the onset of counterion release.

We wish to point out two features of the behavior at small $d$.
The first is that in our model a diverging repulsive pressure is obtained in
the limit of vanishing $d$  because of the assumption of fixed surface
charge. However, effects such as charge regulation \cite{paresgian2} and lipid
demixing \cite{grunberg2,haim1,maror} can modify this assumption and may lead to
to an overall attractive pressure for any $d$.

The second is that attractive van der Waals (vdW) interactions
always prevail at small separations, $d<2$\,nm (in addition to the
electrostatic interactions considered here). In experiments for
small enough $d$, this vdW interaction is stronger than the
electrostatic one and will overcome its repulsion. Thus, the only
way to observe the pure electrostatic crossover between attractive
and repulsive interactions
is to work in a set-up where this crossover occurs for a $d$ range
beyond the influence of vdW interactions.

\section{Acknowledgements}
We thank N. Kampf, J. Klein and S. Perkin for helpful discussion
and for sharing with us their unpublished experimental results. One of us (DA)
acknowledges support from the Israel Science Foundation
(ISF) under grant no.
160/05, the US-Israel Binational Foundation (BSF) under grant no.
287/02, and a Weston visiting professorship award at Weizmann Institute where the latest
part of this work was completed.  SAS is grateful to the BSF and an EU
Network grant for their support.


\end{document}